\documentclass[conference]{IEEEtran}
\usepackage{blindtext, graphicx}
\usepackage[font=small,skip=0pt]{caption}

\ifCLASSINFOpdf
\else
\fi
\begin{document}
%\title{Security and Trust in SOA-based architecture for the Internet of Things}
\title{Trustworthy and Secure Service-Oriented Architecture for the Internet of Things}

\author{\IEEEauthorblockN{Ahmad-Atamli Reineh \\ atamli@cs.ox.ac.uk \\ University of Oxford}
\IEEEauthorblockA{}
\and
\IEEEauthorblockN{Andrew J. Paverd \\ andrew.paverd@cs.ox.ac.uk \\ University of Oxford}
\IEEEauthorblockA{}
\and
\IEEEauthorblockN{Andrew P. Martin \\ andrew.martin@cs.ox.ac.uk \\ University of Oxford}
\IEEEauthorblockA{}}


%Telephone: (800) 555--1212, Fax: (888) 555--1212}


\maketitle

%\begin{abstract}
%\boldmath
%\blindtext[1]
%\end{abstract}

%begin{IEEEkeywords}
%Internet of things, Trust, Security, Middleware, SOA.
%\end{IEEEkeywords}

% \fi
\IEEEpeerreviewmaketitle

\section{Introduction}
In the Internet of Things (IoT), heterogeneous devices connect to each other and to external systems to exchange data and provide services.
%Communication can be based on a client-server paradigm or a peer-to-peer model depending on the function of the IoT system.
Given the diversity of devices, it is becoming increasingly common to establish collaborative relationships between devices to provide composite services.
%For example, in the smart home, the intelligent distributed temperature sensors could collaborate with the smart heating system to automatically regulate the temperature throughout the building. 
However, due to the high degree of heterogeneity in the IoT context, one of the most significant challenges is to develop software applications that can run on a wide variety of devices and can communicate and collaborate with an even wider array of systems.
A common middleware infrastructure for these devices will therefore have a significant impact on the design, deployment, and use of services in IoT systems by allowing developers to focus on the applications rather than the low-level implementation details each device.
%It will also help to reduce the gap between traditional IT systems and IoT devices real world.
%The possibility of a common middleware architecture for IoT has recently gained the attention of researchers due to its potential for simplifying the development of new services and the integration of new different technologies. 
%Such an infrastructure would enable programmers to focus on developing applications without having to deal with low-level details of specific devices and different technologies. 
Spiess et al. \cite{Spiess2009} have proposed and demonstrated a middleware architecture for IoT systems based on the Service-Oriented Architecture (SOA) paradigm which is well suited to supporting heterogeneous devices and communication technologies.
%This paradigm also simplifies the associations between objects and provides coordination between different services. 
%Their architecture consists of multiple layers using principles similar to those explained by de Deugd et al. \cite{Deugd2006}. 
Their architecture uses multiple layers to hide the underlying complexity and heterogeneity of the IoT hardware, software, and protocols
All communication takes place via standardized web-services.

In their architecture, Spiess et al. \cite{Spiess2009} define a single \emph{security layer} and provide a brief description of the intended functionality of this layer.
However, we argue that security is a key requirement of any IoT system and that there are significant benefits of integrating security throughout the whole middleware architecture.
Furthermore, we suggest that the concept of \emph{trust} is also critical to the widespread adoption of this type of middleware for IoT systems.
In order to achieve these objectives, we propose an enhanced version of the middleware architecture by Spiess et al. \cite{Spiess2009} in which security and trust are integrated throughout the architecture, rather than being confined to a single layer.

\section{Security and Trust in IoT}
Security is critical for almost every IoT system, and in the last few years we have witnessed many stark examples of security breaches in smart devices \cite{Best2014}\cite{Parmar2012}. 
%Specifically in the IoT context, there are various threats against which our devices and middleware must be protected. 
Policy-based security models are a well-understood means of enforcing security policies within systems. 
For example, modern mobile devices running Android or iOS use policy-based security models for all applications.
In order to manage the wide range of services available in the IoT context, our architecture provides a similar type of security model for IoT devices. 
Similarly to the mobile devices, our security model ensures that an application is only able to access device features and services for which it has permission and thus cannot interfere with other applications.

Our architecture also provides the property of trustworthiness, which we view as a complementary property to security:
on the one hand, security aims to ensure that the middleware and the underlying devices are protected from malicious applications and external adversaries.
On the other hand, these devices and the middleware must provide suitable guarantees of their trustworthiness to both applications and users before they can be used in certain types of systems, especially those in which there are privacy or safety considerations.
We provide a mechanism through which applications can establish the trustworthiness of the middleware and verify its properties before using it.
In this paper we focus specifically on the security and trust mechanisms for applications and middleware.
Although there are further considerations with regard to security and trust in individual devices, these are left as future work.

%We present our latest findings of the security and trust mechanisms in the mentioned architecture and explain the need for distribution of these features in different layers.

%In order to gain the trust of the public in the Internet of Things, we need to ensure the same failures with respect to security do not come to pass with this system, by ensuring the appropriate mechanisms to guarantee such things exist from the onset of this promising technology.

\section{Enhanced Architecture}

\begin{figure*}
\centering
\includegraphics[scale=0.725]{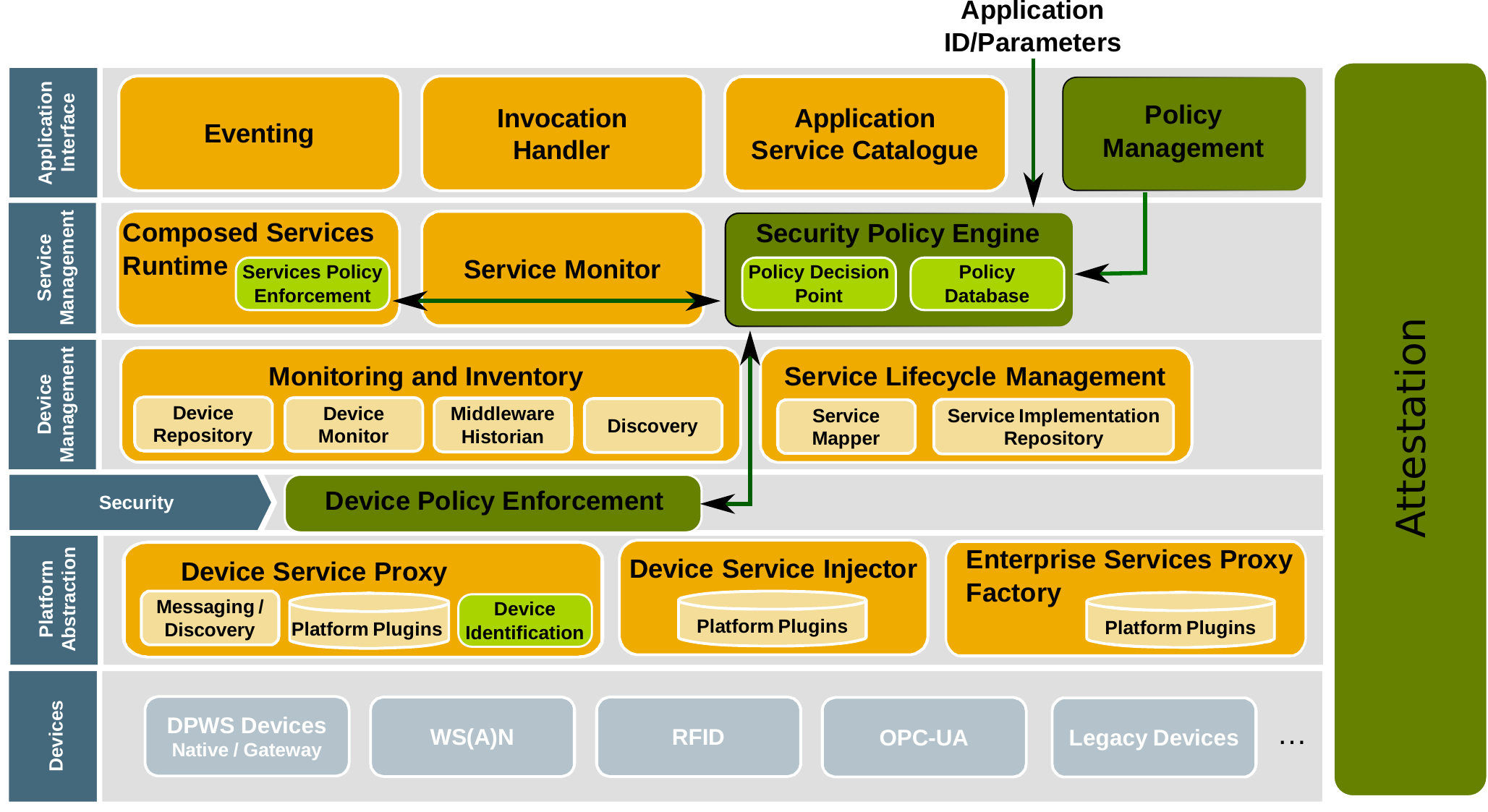}
\setlength{\abovecaptionskip}{0mm}
\caption{Enhanced Middleware Architecture (adapted from \cite{Spiess2009})}
\label{fig:architecture}
\end{figure*}

Figure~\ref{fig:architecture} shows our enhanced architecture as an adaptation of the architecture presented by Spiess et al. \cite{Spiess2009}.  
In their architecture security is presented in one layer whereas we show how security and trust mechanisms should be integrated throughout the architecture. 
The original architecture is shown in yellow with our enhancements presented in green.

\subsection{Security}

\textbf{Policy Management:} 
The policy management component facilitates adding, removing and modifying policies in the system. 
The owner or user of the system can manage the policies through this interface. 
The implementation of this interface is specific to each system.

\textbf{Security Policy Engine:} 
The security policy engine is the main policy component of the system.
It contains the system-wide policy database in which the security rules are stored.
It also contains the central policy decision point (PDP). 
The PDP receives information about the application (e.g. application ID, signature and parameters) and uses this to make policy decisions based on the rules in the database. 
These decisions are passed to the relevant policy enforcement points.

\textbf{Device Policy Enforcement (DPE):} 
This component is responsible for enforcing the decisions from the PDP whenever an application is dealing with a specific device. 
For example, Spiess et al. \cite{Spiess2009} have demonstrated how their architecture could be used on a robotic arm. 
However, for safety reasons, a sensing application may only have permission to read the position of the arm but not to move it.
The DPE would allow access to the position data but deny any requests from that application to move the arm.

\textbf{Service Policy Enforcement (SPE):} This component performs a similar function to the DPE component but deals with composite services rather than individual device capabilities.
For example, the security policy could mandate that any device which has access to sensitive information must not have direct access to the Internet.
This type of service-level policy would be enforced by the SPE component.

\subsection{Trust}

\textbf{Device Identification:} 
The device identification component is responsible for identifying devices in the system to guarantee that only legitimate devices are used by the middleware and applications.

\textbf{Attestation:} 
At each layer in the architecture, attestation is used to provide a guarantee to the application that the middleware it is using is trusted. 
Theoretically, each layer in the middleware can be hosted in a different location, but from a security point of view, this makes it harder to ensure that all hosting devices are trusted. 
The application would like a guarantee that the middleware hosted in all devices is trusted and thus attestation is necessary.
In order to hide the lower layers from the developer, each component in the middleware should be able to make trust decision about the components with which it communicates and communicate this decision up the stack to the Application Interface.

\section{Examples}
We present two hypothetical scenarios, based on the demonstration by Spiess et al. \cite{Spiess2009}, to illustrate the necessity of our security and trust enhancements.
In their demonstration system, a robotic arm is operating a manufacturing process and a sensor is checking that the vibration is within acceptable limits.
As a hypothetical extension, consider a web-based efficiency monitoring application that uses the middleware to monitor the position of the robotic arm.
Without fine-grained device policy enforcement, this application might not only have access to the position of the arm but might also be able to move it.
Since it is Internet-connected, this application might become compromised.
This supposed monitoring application could then be used in a denial of service attack by introducing small vibrations in the arm at exactly the right frequency to trigger the vibration sensor and halt the process.
Our enhancements would mitigate this in two ways: 1) fine-grained policy enforcement at the DPE would prevent the monitoring application from moving the arm and 2) the SPE could enforce mutual exclusion between Internet-connectivity and permission to control the robotic arm. 
As an example of the need for trust in the middleware and devices, consider the emergency stop button.
The application expects that this signal is always a real event and therefore stops the process.
However, untrusted middleware could falsify this signal causing the process to stop for no reason.
Due to safety concerns, ignoring this signal is not an option.
Our enhancements would mitigate this attack by using strong device identities and attestation so that the application can always make an informed decision based on the trustworthiness of the underlying system.

%\section{Conclusion and Future work}
%The SOA middleware for IoT systems proposed by Spiess et al. simplifies the development of new services and the integration of new technologies. 
%However, it does not go into detail about the security considerations which are all placed in a single layer. 
%We argue that security can not reside in one layer and has to be distributed among the different layers of the architecture. 
%We present the outline of an enhanced architecture for trustworthy and secure middleware in which the mechanisms for enforcing security and establishing trust are distributed throughout the %architecture.
%As future work, we plan to study security and trust considerations under different adversary models.
%In this work we assumed that the devices are trusted and our adversary model was only concerned with the application and the middleware. 
%However, a malicious device could have serious consequences for the system. 
%In order to guarantee the trustworthiness of the system, we need to consider the trustworthiness of all components.

\raggedbottom

\raggedright
\bibliographystyle{IEEETran}
\bibliography{IoT}
\end{document}